\documentclass[aps,prl,groupedaddress,showkeys,twocolumn,superscriptaddress]{revtex4-1}
\usepackage{graphicx}
\usepackage{dcolumn}
\usepackage{bm}

\makeatletter
\AtBeginDocument{\@ifpackageloaded{natbib}{\ifNAT@numbers\if@filesw\immediate\write\@auxout{\string\global\string\NAT@numberstrue}\fi\fi}{}}
\makeatother
\begin{document}


\title{Interlayer Diffusion of Nuclear Spin Polarization in $\nu=2/3$ Quantum Hall States}


\author{{Minh-Hai Nguyen}}
\email[]{mn455@cornell.edu}
\altaffiliation{Department of Physics, Cornell University, Ithaca, NY 14853, USA}
\affiliation{Department of Physics, Graduate School of Science, Kyoto University, Kyoto 606-8502, Japan}

\author{{Shibun Tsuda}}
\affiliation{Department of Physics, Graduate School of Science, Kyoto University, Kyoto 606-8502, Japan}

\author{Daiju Terasawa}
\author{Akira Fukuda}
\affiliation{Department of Physics, Hyogo College of Medicine, Nishinomiya 663-8501, Japan}

\author{Yangdong Zheng}
\author{Anju Sawada}

\affiliation{Research Center for Low Temperatures and Materials Sciences, Kyoto University, Kyoto 606-8501, Japan}

\date{\today}

\begin{abstract}
At the spin transition point of $\nu = 2/3$ quantum Hall states, nuclear spins in a two-dimensional electron gas are polarized by an electric current. Using GaAs/AlGaAs double-quantum-well samples, we  first observed the spatial diffusion of nuclear spin polarization between the two layers when the nuclear spin polarization is current-induced in one layer. By numerical simulation, we estimated the diffusion constant of the nuclear spin polarization to be $15 \pm 7$\,nm$^2$/s.
\end{abstract}


\maketitle


\label{sec:Intro}

The nuclear spin degree of freedom in semiconductor nanostructures has  attracted a great deal of attention. Because of their very long coherence time, nuclear spins are one possible candidate for the qubits of quantum computing \cite{Kane98, Privman98, Bennett00, Mozyrsky01, Mani02}, but they are likely disturbed in solid-state devices because of their interaction with the neighboring electron system via the hyperfine interaction. In quantum Hall (QH) systems, however, the quantization of electrons’ energy imposes a strong restriction on energy conservation. The Zeeman energy of electrons, which is about 3 orders of magnitude larger than that of nuclei, must be compensated for. As a result, methods to control nuclear spins in QH systems have long been confronted with many physical and technical difficulties.

At the spin transition point of $\nu = 2/3$ quantum Hall states (QHSs), the spin up and spin down composite fermions' Landau levels degenerate energetically, allowing the flip-flop exchange of nuclear spins with electron spins through the hyperfine interaction. This opens up the opportunity to interact with the nuclear spins system via electrical means. In fact, for these QHSs, hysteretic transport and anomalous magnetoresistance peaks have been observed \cite{Kumada02, Kumada04, Kronmuller98, Kronmuller99, Smet01, Smet02, Hashimoto02, Kraus02, Stern04}. Resistively detected nuclear magnetic resonance measurements have shown the involvement of  dynamic nuclear spin polarization (DNP) in originating the magnetoresistance peaks \cite{Kronmuller98, Kronmuller99, Hashimoto02, Kraus02, Stern04}. However, how DNP connects with magnetoresistance remains an open question. It has been reported that DNP carried out by an electric current varies proportionally to the magnetoresistance \cite{Hashimoto02}, but the physical mechanism is still under debate.

To control nuclear spins and explain their relation to the anomalous magnetoresistance peaks, it is necessary to understand the nuclear spin dynamics in QH systems. Because nuclear spin polarization can diffuse through a dipole-dipole interaction, it is possible for the magnetoresistance of one layer in a bilayer QH system to be affected by DNP in the other layer. In this Letter, we perform transport measurements at the spin transition points of $\nu = 2/3$ QHSs using a GaAs/AlGaAs bilayer sample to investigate how current-induced DNP of a layer propagates to the other layer by measuring the temporal evolution of the magnetoresistances of both layers.

\label{sec: Experiments}

The sample, a modulation-doped double quantum well grown by molecular beam epitaxy, consists of two 20-nm-wide GaAs wells separated by a 1.5-nm-thick AlAs barrier.  The tunneling gap $\Delta_{\mathrm{SAS}}$ for this sample is calculated to be about 5\,K. The low-temperature mobility is 220\,m$^2 (\rm Vs)^{-1}$ at a total electron density of $2.0 \times 10^{15}$\,m$^{-2}$. The sample was fabricated by conventional photolithography into a 50-nm-wide Hall bar with a voltage probe distance of 180\,nm. By applying front- and back-gate biases, the electron densities in the front and back layers can be controlled independently. The sample is immersed in the mixing chamber of a dilution refrigerator with a 62\,mK base temperature. A static magnetic field of 6\,T generated by a superconducting magnet is applied perpendicular to the two-dimensional electron gas (2DEG) plane. We measure magnetoresistances by a standard low-frequency AC lock-in technique with a reference frequency of 17.7\,Hz.

\label{sec:Results}

To confirm the spatial propagation of DNP, in our first experiment, we pump DNP in one layer and measure the magnetoresistance $R$ in the other layer. After the initial magnetoresistances $R_0$ of the two layers are measured, the front layer is set at the spin transition point of the $\nu=2/3$ QHS while electrons in the back layer are depleted. A current $I$ is pumped through the front layer for 60\,min, which is long enough for the DNP in the front layer to saturate. Then the saturated magnetoresistances $R_{\mathrm{sat}}$ of the front and back layers are measured. When measuring $R$ of a layer, we set that layer at the $\nu=2/3$ QHS and deplete electrons in the other layer, then we measure $R$ for a short time  (4\,s) by using a current as low as 10\,nA (so that the measurement does not noticeably affect the DNP).

\begin{figure}
\includegraphics[scale=0.35]{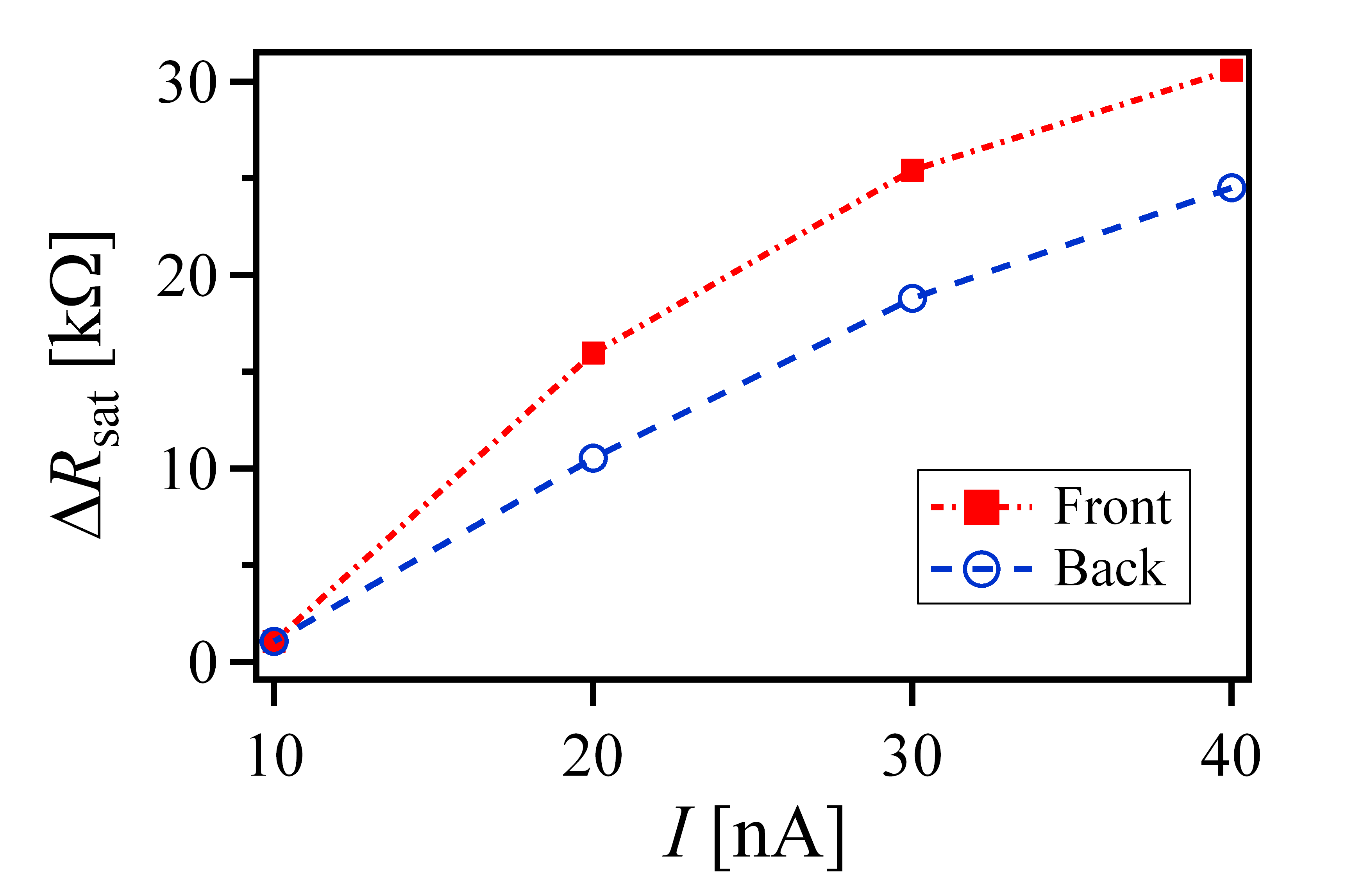}
\caption{(Color online). Plot of saturated magnetoresistance enhancements $\Delta R_{\mathrm{sat}}$ versus pumping currents. DNP in the front layer is pumped by currents $I$ for 60 min so that the magnetoresistances $R$ of both layers increase to saturated values. For currents less than 10\,nA, a negligible change of $R$ is observed. For higher currents, $\Delta R_{\mathrm{sat}} = R_{\mathrm{sat}} - R_0$ of the front layer (solid squares, red) increases with the current, while that of the back layer (open circles, blue) also increases although there is no current flowing in the back layer.}
\label{fig:Current}
\end{figure}

The saturated magnetoresistance enhancements $\Delta R_{\mathrm{sat}} = R_{\mathrm{sat}} - R_0$ of the front and back layers (which are proportional to the saturated nuclear spin polarizations in the respective layers) at different pumping currents are plotted in Fig.\,\ref{fig:Current}. The $I=10$\,nA current does not considerably pump DNPs in both the front and back layers. For $I > 10$\,nA, $\Delta R_{\mathrm{sat}}$ of the front layer increases with $I$. This means that, within the range of $I = 10$--40\,nA, DNP in the front layer increases with the flow of charge carriers across the layer, which is consistent with previous measurement \cite{Kronmuller98}. Surprisingly, nonzero values of $\Delta R_{\mathrm{sat}}$ of the back layer are observed although electrons in the back layer are depleted. Notice that $\Delta R_{\mathrm{sat}}$ of the back layer also increases with the pumping current flowing in the front layer,  strongly indicating that current-induced DNP in the front layer propagates to the back layer.

\begin{figure}
\includegraphics[scale=0.20]{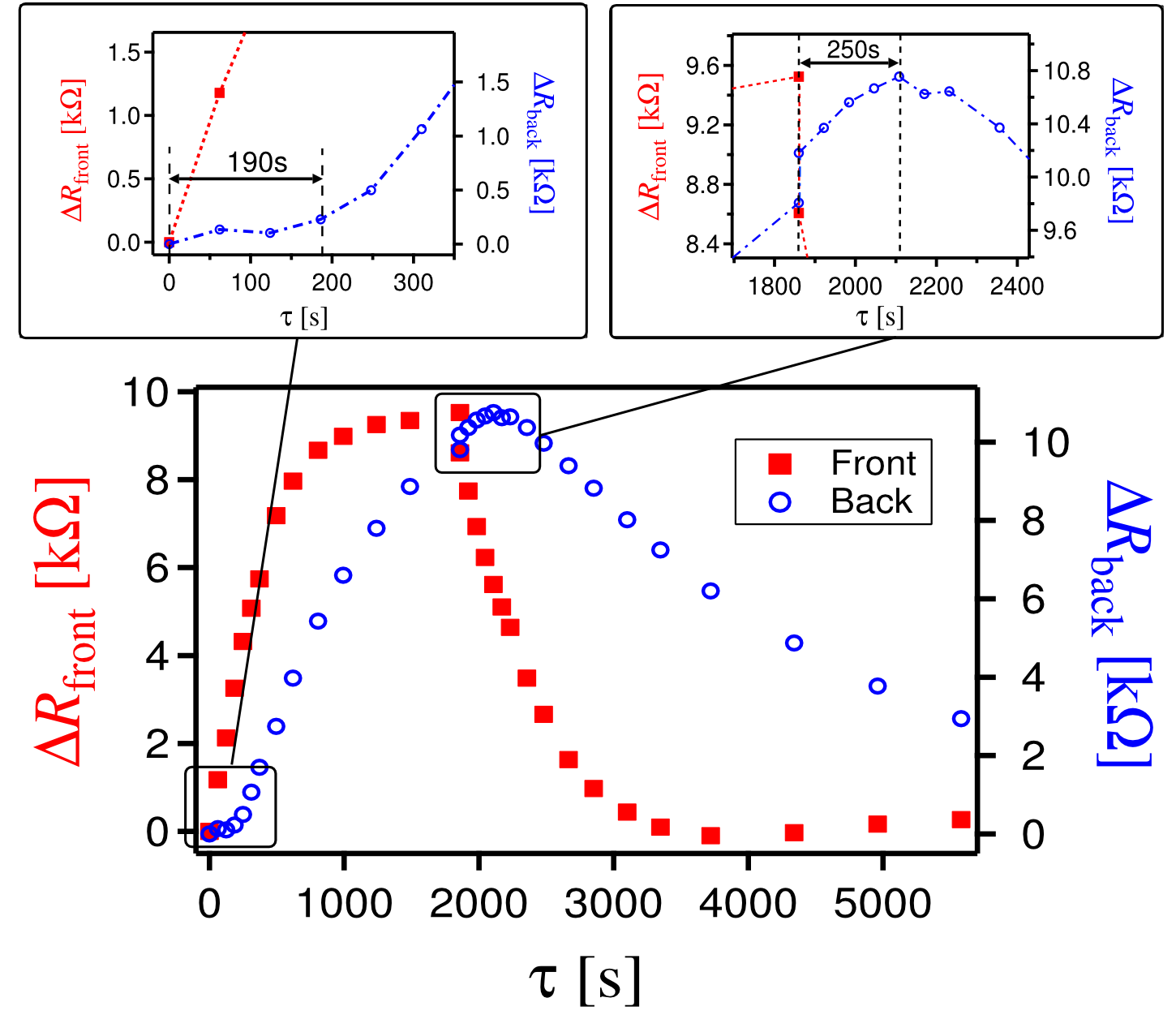}
\caption{(Color online). Time evolution of the magnetoresistance enhancements of the front and back layers. In the first 1,860\,s, DNP in the front layer is pumped by a 20-nA current. The magnetoresistance enhancement of the front layer, $\Delta R_{\mathrm{front}}$ (solid squares, red, left axis), increases immediately and  quickly approaches a saturated value. However, that of the back layer, $\Delta R_{\mathrm{back}}$ (open circles, blue, right axis), increases more slowly after a delay of about 190\,s. At time $\tau = 1,860$\,s, the current is set to zero. $\Delta R_{\mathrm{front}}$ drops immediately and quickly saturates  to the original value. However, $\Delta R_{\mathrm{back}}$ exhibits a delay of about 250\,s and a slow decrease. The enlarged figures in the insets illustrate the delayed response of $\Delta R_{\mathrm{back}}$ to $\Delta R_{\mathrm{front}}$.}
\label{fig:UpDown}
\end{figure}

To investigate the dynamics of the DNP propagation, we perform a temporal measurement of DNPs in the two layers. First, we set the front layer at the spin transition point of the $\nu=2/3$ QHS and deplete electrons in the back layer. A current of 20\,nA is sent through the front layer to pump DNP. At time $\tau$ we measure $R$ of the two layers (using the same measuring procedure as in the first experiment). In this way we can observe how $R$ (and therefore DNP) in the two layers varies with time. At $\tau = 1,860$\,s, we turn off the pumping current.

Figure\,\ref{fig:UpDown} shows the time evolution of the magnetoresistance enhancements of the front and back layers. In the first 1,860\,s, after a 20-nA current is sent through the front layer, its magnetoresistance enhancement $\Delta R_{\mathrm{front}}$ (solid squares, red, left axis) rises immediately and approaches a saturated value. However, that of the back layer, $\Delta R_{\mathrm{back}}$ (open circles, blue, right axis), exhibits a delay of about 190\,s (left inset) before gradually increasing at a rate lower  than $\Delta R_{\mathrm{front}}$. At time $\tau=1,860$\,s, the pumping current is set to zero but the states of the two layers are kept unchanged. $\Delta R_{\mathrm{front}}$ drops immediately and rapidly to zero. This is not surprising since the fluctuation of the conducting charge carriers in the front layer diminishes the DNP. What intrigues us is the delayed behavior of $\Delta R_{\mathrm{back}}$. When $\Delta R_{\mathrm{front}}$ starts to drop, $\Delta R_{\mathrm{back}}$ continues to rise and attains its peak  $\sim$250\,s later (right inset) before beginning to fall more slowly than $\Delta R_{\mathrm{front}}$. It is clear that the propagation of the current-induced DNP from the front layer to the back layer is due to spatial nuclear spin diffusion.

\begin{figure}
\includegraphics[scale=0.18]{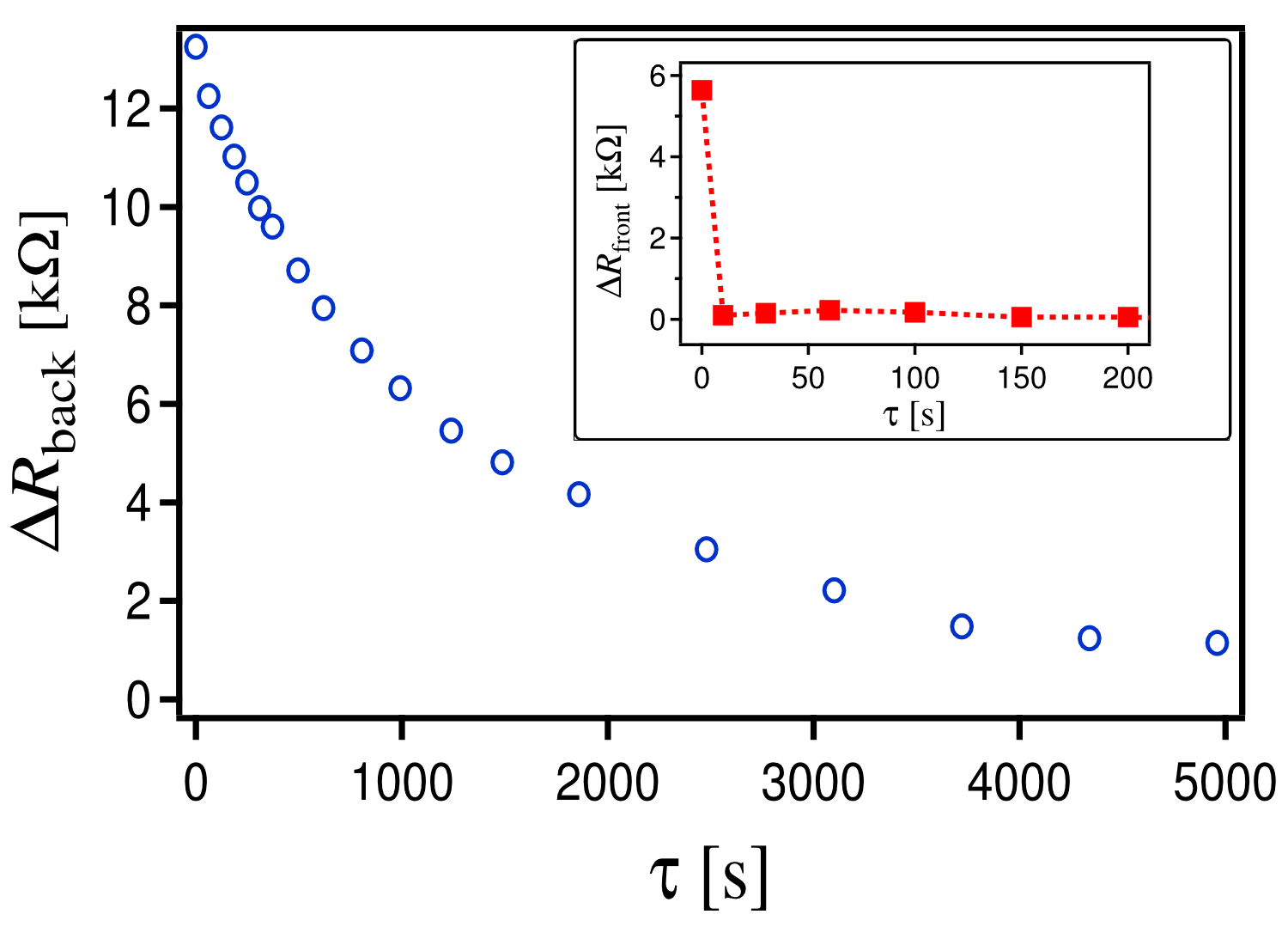}
\caption{(Color online). Time evolution of the magnetoresistance enhancement of the back layer after the front layer is set in a $\nu=1.15$ skyrmion lattice state. The magnetoresistance enhancement of the back layer decreases slowly toward zero with a time constant of 1,900\,s. (Inset) DNP in the front layer, showing how it disappears completely after 10\,s.}
\label{fig:Skyrmion}
\end{figure}

To reaffirm the existence of interlayer DNP diffusion, we perform another experiment which is also used to estimate the nuclear spin diffusion constant later in this Letter. We first pump DNP in the front layer by using a current of 20\,nA for 14\,h, which is a sufficiently long time for the DNP to propagate throughout the sample. Then we turn off the pumping current. While the back layer is kept in the depletion state, the front layer is set in the $\nu = 1.15$ skyrmion lattice state \cite{Brey95} in which nuclear spins are allowed to relax quickly toward equilibrium by the Goldstone mode of skyrmions \cite{Cote97, Hashimoto02}. At time $\tau$ we measure $R$ of the two layers. We confirm that DNP in the front layer vanishes completely in less than 10\,s (see Fig.\,\ref{fig:Skyrmion} inset). That of the back layer, however, decreases very slowly, as shown in Fig.\,\ref{fig:Skyrmion}, with a time constant of 1,900\,s. This decay of $\Delta R_{\mathrm{back}}$ can be attributed to two factors: the spin-lattice interaction and  DNP diffusion to the front layer.

\begin{figure}
\includegraphics[scale=0.38]{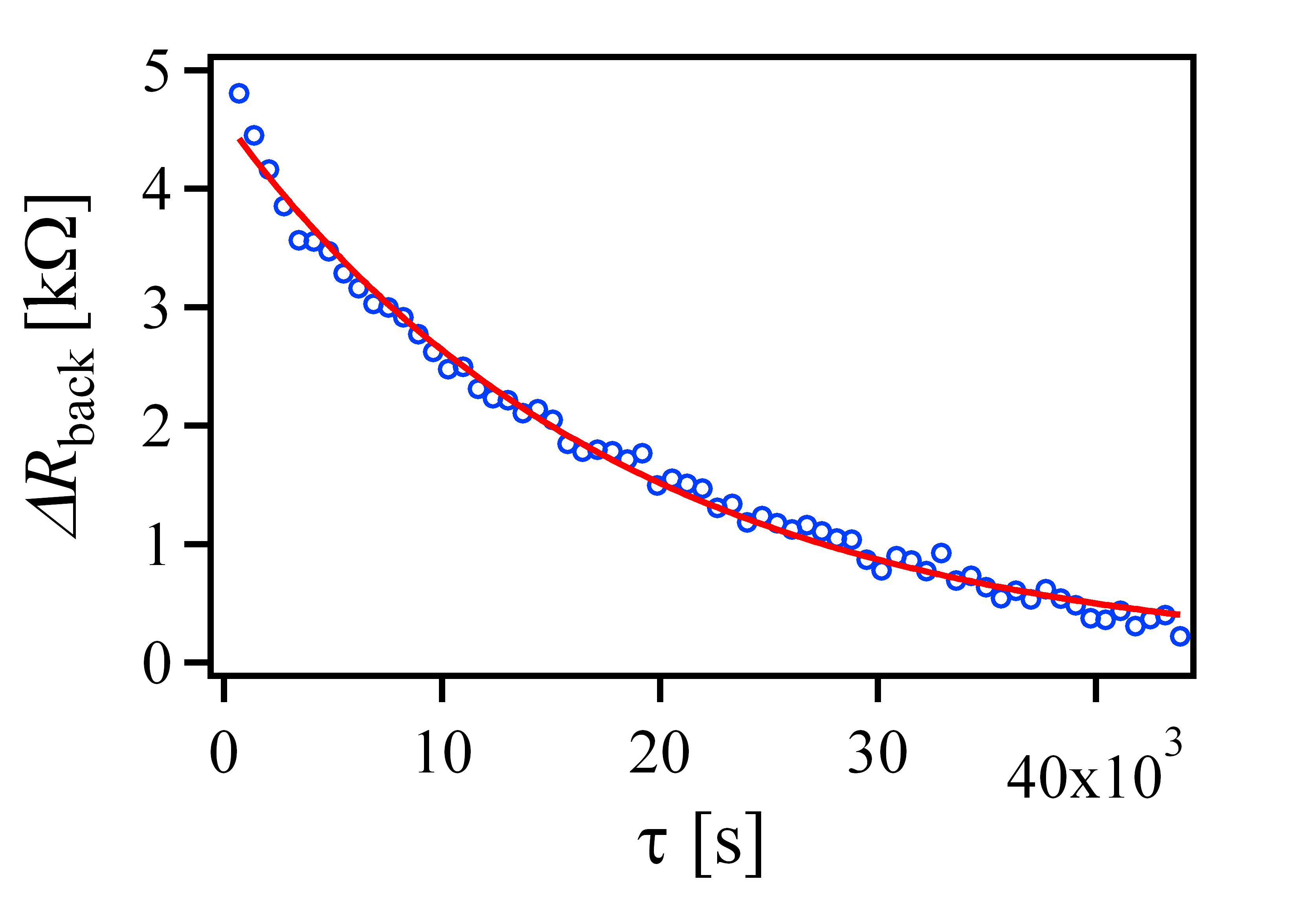}
\caption{(Color online). Time evolution of the magnetoresistance enhancement of the back layer, $\Delta R_{\mathrm{back}}$, after electrons in both layers are depleted. Without charge carriers, $\Delta R_{\mathrm{back}}$ decays very slowly because of spin-lattice interaction. The solid line shows the exponential fit. The time constant is $T_{d\,0} = 18 \times 10^3$\,s.}
\label{fig:Decay}
\end{figure}

To determine which factor is essential to the decay of $\Delta R_{\mathrm{back}}$, we measure the effect of the spin-lattice interaction only. DNP in the front layer is pumped for a sufficiently long time. By doing so, we can rule out the effect of DNP spatial diffusion. Then we deplete electrons in both layers. The measured magnetoresistance enhancement of the back layer is shown in Fig. \ref{fig:Decay}. Without charge carriers in both layers, DNP in the back layer decays extremely slowly with a time constant of $T_{d\,0} = 18 \times 10^3$\,s, about 10 times longer than that of Fig.\,\ref{fig:Skyrmion}. It is obvious that the effect of spin-lattice interaction is very small and the decay of $\Delta R_{\mathrm{back}}$ in Fig.\,\ref{fig:Skyrmion} is essentially caused by the DNP interlayer diffusion.

\label{sec:Simulation}

We performed a numerical simulation of the DNP diffusion process to estimate the diffusion constant. Consider the  DNP dynamics in the experiment shown in Fig.\,\ref{fig:UpDown}. In the front layer, DNP is pumped by a current for the first 1,860\,s, diffuses in the $z$ direction (perpendicular to the 2DEG plane), and decays owing to interaction with fluctuating charged carriers and the spin-lattice interaction.

In the current-induced pumping process, DNP is shown experimentally to vary exponentially with time \cite{Hashimoto02} and roughly linearly with the current density (as shown in Fig.\,\ref{fig:Current}). The nuclear spin polarization $P(z,\tau)$ can be approximately described as
\begin{equation}
\frac{\partial P(z,\tau)}{\partial \tau} = \frac{P_{\mathrm{sat}} - P(z,\tau)}{T_p} | \Psi(z) |^2
\label{eqn:Pump}
,\end{equation}
where $P_{\mathrm{sat}}$ is the saturated DNP, $1/T_p$ is the pumping rate, and $\Psi(z)$ is the wave function of electrons confined in the infinite square quantum well respective to one layer.

Similarly, the decay of DNP in the $\nu=2/3$ QHSs owing to fluctuating electrons also varies exponentially with time and is proportional to the density of conducting electrons. Thus it can be given by
\begin{equation}
\frac{\partial P(z,\tau)}{\partial \tau} = - \frac{P(z,\tau)}{T_{d\,2/3}} |\Psi(z) |^2
\label{eqn:Decay1}
,\end{equation}
where $T_{d\,2/3}$ is the time constant of the relaxation in the $\nu=2/3$ QHSs. In the depletion state, the decay of DNP owing to the spin-lattice interaction can be described as
\begin{equation}
\frac{\partial P(z,\tau)}{\partial \tau} = - \frac{P(z,\tau)}{T_{d\,0}}
\label{eqn:Decay2}
,\end{equation}
where $T_{d\,0}=18 \times 10^3$\,s is the time constant of the relaxation caused by spin-lattice interaction.

If we assume that DNP diffuses only in the $z$ direction, the diffusion process of DNP can be simply described by the one-dimensional diffusion equation
\begin{equation}
\frac{\partial P(z,\tau)}{\partial \tau} =  D \frac{\partial ^2}{\partial z^2} P(z,\tau)
\label{eqn:Diffusion}
,\end{equation}
where $D$ is the diffusion constant. The combination of Eqs.\,(\ref{eqn:Pump})--(\ref{eqn:Diffusion}) describes the dynamics of DNP in the front layer in the $\nu=2/3$ QHS during the first 1,860\,s. For $\tau > 1,860$\,s, the pumping current is turned off, so Eq.\,(\ref{eqn:Pump}) is ignored. In the back layer, which is depleted, only the diffusion process [Eq.\,(\ref{eqn:Diffusion})] and the relaxation owing to the spin-lattice interaction [Eq.\,(\ref{eqn:Decay2})] take place.

The relation between total $P(z,\tau)$ and $\Delta R$ is observed to be linear \cite{Hashimoto02}, i.e.,

\[\Delta R_{\mathrm{front}}(\tau) = C_f \int_{\mathrm{front}} P(z,\tau)dz,\]
\begin{equation}
\Delta R_{\mathrm{back}}(\tau) = C_b \int_{\mathrm{back}} P(z,\tau)dz,
\end{equation}
where the integrals are taken over the thicknesses of the layers and $C_f$ and $C_b$ are constants. Using this linear relation, we fit our normalized simulation values to normalized measured $\Delta R$.

\begin{figure}
\begin{flushleft}
(a)\\
\includegraphics[scale=0.36]{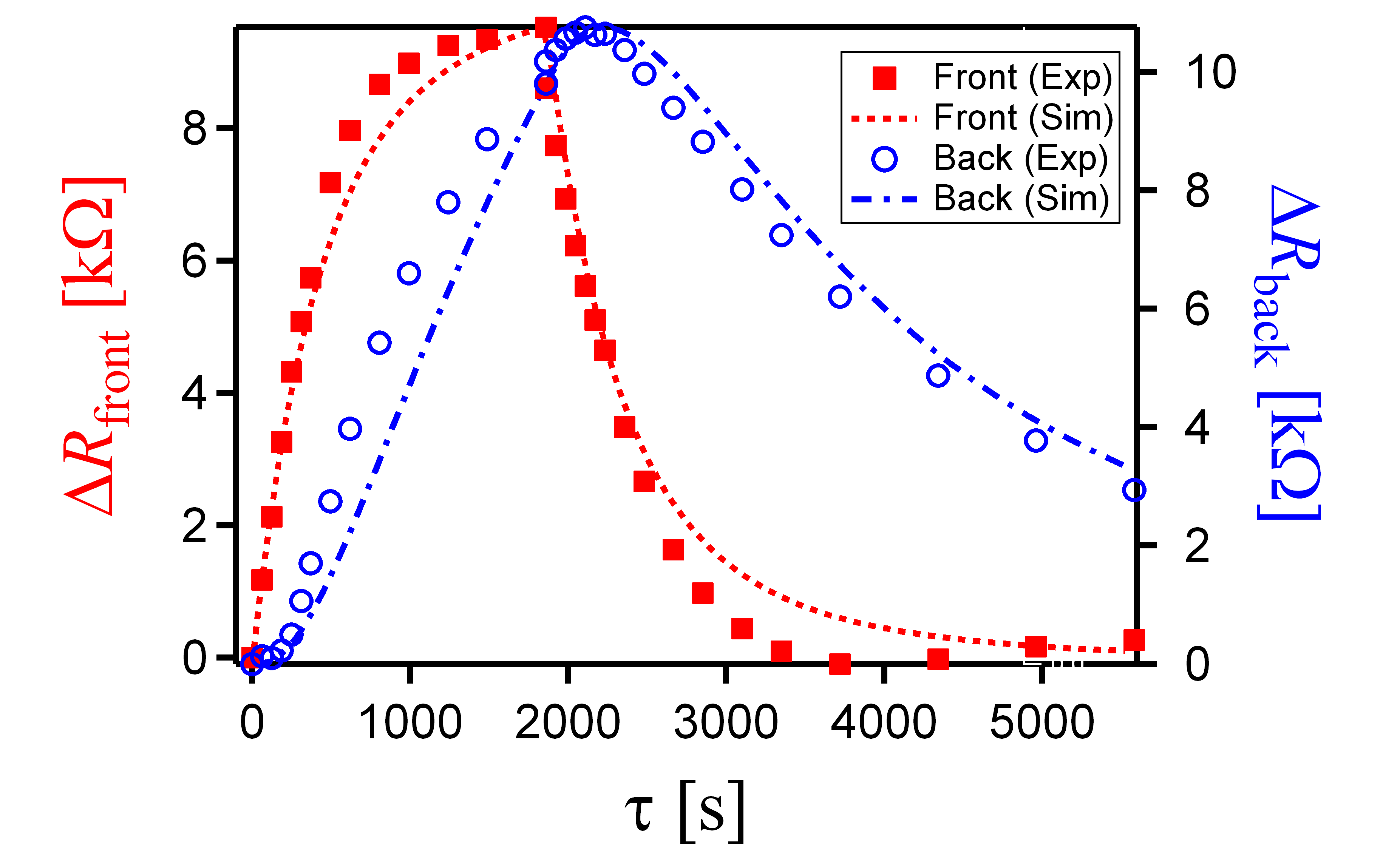}
(b)\\
\includegraphics[scale=0.32]{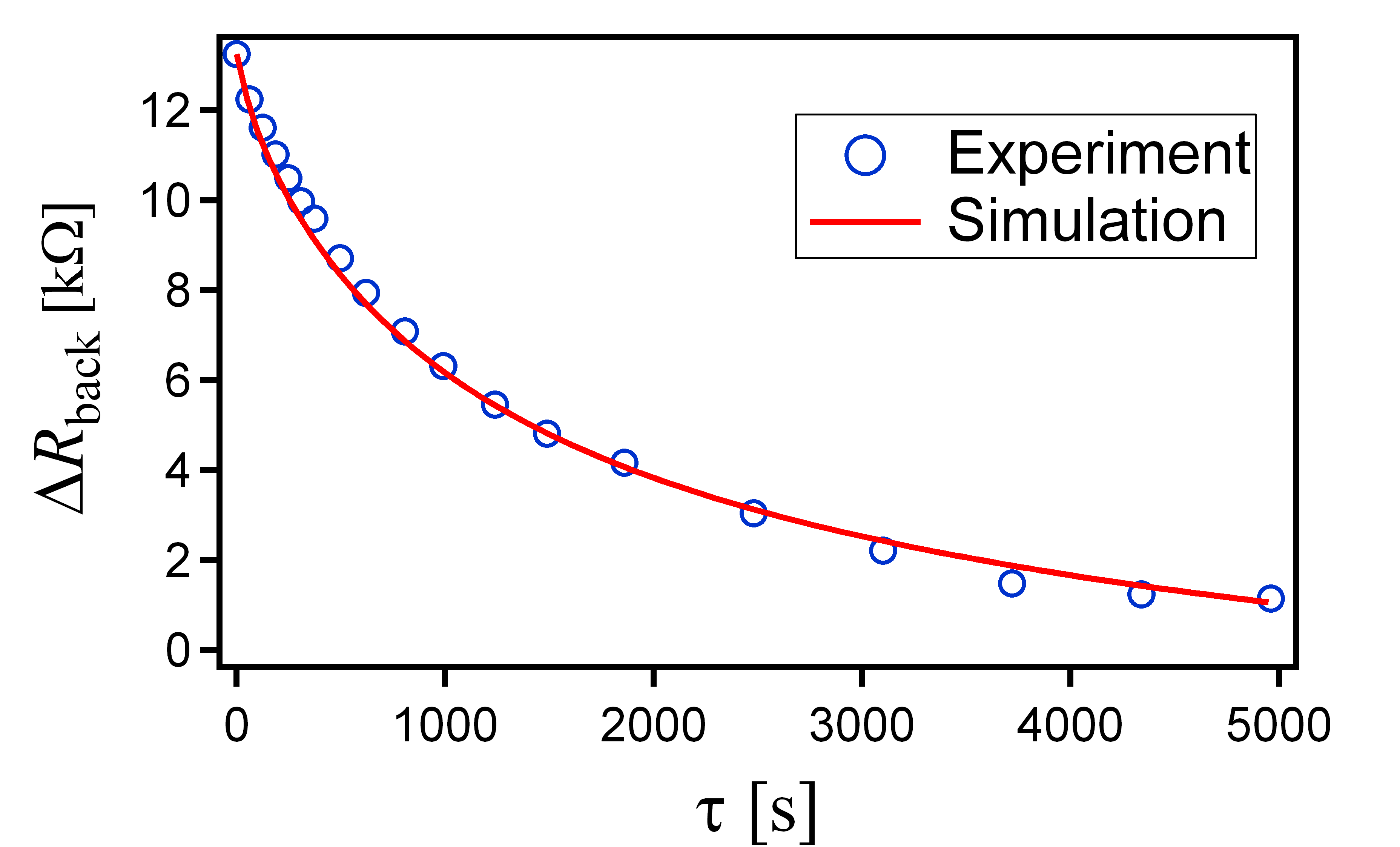}
\end{flushleft}

\caption{(Color online). Simulation results (lines) fitted to measured magnetoresistance enhancements (squares and circles). (a) Fit to data of Fig.\,\ref{fig:UpDown};  estimated values of the parameters are $D = 20\;\mathrm{nm}^2$/s and $T_p = 10$\,s. (b) Fit to data of Fig.\,\ref{fig:Skyrmion}; estimated $D = 8\;\mathrm{nm}^2$/s.}
\label{fig:Sim}
\end{figure}

The simulation results are shown in Fig.\,\ref{fig:Sim}(a). By fitting these results to measured data, the parameters are estimated to be  $D = 20$\,nm$^2$/s and $T_p = 10$\,s. We also performed simulation and fitting to the experimental data of Fig.\,\ref{fig:Skyrmion}. Here DNP in the front layer is always zero and serves as the Dirichlet boundary condition. In the back layer,  diffusion and decay processes occur. The fitting result is shown in Fig.\,\ref{fig:Sim}(b) and the estimated diffusion constant is $D = 8$\,nm$^2$/s.

We repeat the experiments with the roles of the layers exchanged (i.e., we pump DNP in the back layer and measure its propagation to the front layer) and obtain consistent results. The overall estimated parameters are $D = 15 \pm 7$\,nm$^2$/s and $T_p = 7 \pm 3$\,s (average values of repeated measurements and fittings). Our estimated value of the diffusion constant $D$ is consistent with the one obtained by optical measurement for bulk GaAs, which is $10 \pm 5$\,nm$^2$/s \cite{Paget82}. A rough estimation similar to that in \cite{Hayashi08} shows that the diffusion time $t_{sd}$ for DNP to transfer a distance $L = a_0/\sqrt{2}$ (where $a_0 = 0.565$\,nm is the GaAs lattice constant) between atoms of the same element is $t_{sd} \approx {L^2}/{D} = 11\;\mathrm{ms}$, much shorter than the DNP pumping relaxation time $T_p = 7 \pm 3$\,s. This indicates a rapid nuclear spin diffusion regime for GaAs structures. It is evident that the small nuclear spin-lattice relaxation rate $1/T_1 = 10^{-3}$\,s$^{-1}$ \cite{Hashimoto02, McNeil76} is mainly due to the rapid DNP diffusion.

\label{sec:Discusstion}

In our simulation, we used a one-dimensional diffusion model and assumed that DNP diffuses  only in the $z$ direction perpendicular to the 2DEG plane. If DNP is formed along domain walls (according to the electronic spin domains model described in \cite{Kraus02, Stern04}), two- or three-dimensional diffusion model should be used. However, our recent work \cite{Tsuda13} shows that at the spin transition point of the $\nu=2/3$ QHSs, the 2DEG system exhibits an insulation phase. Thus the conducting charge carriers are forced to flow into the 2DEG plane. Assumed that DNP is formed inside the plane of the layers in sufficiently large continuous areas, the in-plane diffusion of DNP is negligible and therefore our one-dimensional diffusion model is plausible.

Although we demonstrate the interlayer diffusion of DNP, how DNP in one layer affects the magnetoresistance of the other layer remains an open question. Within the electronic spin domains model, the diffused DNP from the front layer adds disorder to the domain structure of the back layer and affects its magnetoresistance. Some disorder caused by the diffused DNP complicates the domain wall structure, but some can simplify it. The overall effect is expected to be insignificant. However, our experimental results show that the effect of DNP diffusion is clear and consistent. Therefore, this question requires additional investigation and when answered will shed light on the mechanism of current-induced DNP.

\label{sec:Conclusion}

In conclusion, we have demonstrated the spatial diffusion of DNP between two layers of a bilayer QH system. From the delayed and slow response of the DNP of one layer after DNP is pumped in the other layer, we estimate the DNP interlayer diffusion constant to be $15 \pm 7$\,nm$^2$/s by numerical simulation. This result helps us to better understand DNP dynamics, provides a method to control nuclear spins, and  hints at an explanation for the mechanism of current-induced DNP and its relation to  magnetoresistance.

\begin{acknowledgments}
We thank N.\,Kumada and K.\,Muraki of the NTT Basic Research Laboratories  and  T.\,Hatano and Y.\,Hirayama of Tohoku University not only for providing us  with precious high-mobility samples but also for fruitful discussions. We are also grateful to T.\,Saku for growing the sample heterostructures. This research was supported in part by Grants-in-Aid for Scientific Research (Nos.\,24540319, 24540331, and 25103722).
\end{acknowledgments}

\bibliography{MinhHai_DNP_Diffusion_Ref}

\end{document}